\documentclass[14pts]{IEEEtran}
\usepackage[letterpaper]{geometry}
\begin{document}
\twocolumn[
\begin{@twocolumnfalse}
\title{\Large Northern Hemisphere UHECR data further supports CMBR photons with weak U(1) component}

\author{Daniel Piasecki
\\
\emph{Tulane University, New Orleans, LA 70118}}

\maketitle \thispagestyle{empty}

\begin{abstract}
\normalsize
I further test the theory that the Cosmic Background Radiation (CBR) is made of photons with a weak U(1) component. 
Tipler (2005) has previously argued that the consistency of the Standard Model (SM) with the Second Law of Thermodynamics requires the early universe CBR to be composed mainly of an SU(2) gauge field. The U(1) field would be suppressed. One of the consequences of this approach is that the Ultra High Energy Cosmic Rays (UHECR) would be able to propagate much further than conventionally accepted, as an SU(2) dominant CBR would be largely unable to couple with right handed fermions. I test if this novel theory solves the problem of UHECR origin by finding suitable candidates up to a redshift z = 0.1 within three degrees of the arrival direction. Utilizing the Fly's Eye Northern Hemisphere UHECR data, I identified candidates  with  80\%  success for the Northern Sky UHECR (98.7\% if certain celestial objects which are likely to be Active Galactic Nuclei are absolutely identified as such). This is parsimonious with the CBR theory, which has other important implications for the Standard Model, early universe cosmology, and the origin of matter and anti-matter. This builds off of Tipler and Piasecki (2018), where we have reviewed UHECR data from the Southern Hemisphere to get a 86\% success identification rate. We predict that the remaining UHECR not paired with a potential source will have sources identified upon closer telescopic investigations of these regions. Other recent experiments further suggest an SU(2) dominant composition of the CBR photons. The problem of UHECR origin may be solved. 

\end{abstract}


\end{@twocolumnfalse}]

\section{Introduction and Theory}
One of the greatest unsolved mysteries of current astrophysics is the existence of the Ultra High Energy Cosmic Rays (UHECR). The UHECR are interstellar cosmic rays  which come with such high energy that they were thought to be impossible based on our current understanding of particle interaction with the Cosmic Background Radiation (CBR). We continue to observe their very existence, leading some to suggest the possibility of new physics. 

A very intriguing theory has been proposed by Tipler (2005), which could explain this aberrant astrophysical phenomena using only known principles, without postulating novel entities at play. This idea is based on the well tested Standard Model (SM) of Particle Physics, Second Law of Thermodynamics, and Bekenstein Bound applied to cosmological scales (Bekenstein, 1989). 

The SM reveals that electromagnetic radiation is not fundamental but composed of elementary gauge fields, namely the U(1) and SU(2) gauge fields. From basic principles, Tipler (2005) shows that one of the components of electromagnetism, the U(1) field, could not exist moments after the Big Bang singularity. Thus, the earliest universe could not start out with pure electromagnetism but only with the SU(2) gauge field. This would mean that the CBR photons that came shortly after were mainly composed of an SU(2) gauge field with a very weak or strongly suppressed U(1) field.

We should be able to observe the effects of a strongly SU(2) dominant CBR to this day (Tipler, 2005). Tipler (2005) argues that it is quite likely that the present day CBR would largely retain its SU(2) dominant structure, so the theory is testable inside the laboratory. One of the important properties of an SU(2) dominant CBR is that it does not couple with right handed fermions (such as right handed electrons and quarks). It is suspected that the UHECR could be high energy protons and Iron (Fe) Nuclei. Because the theorized SU(2) dominant CBR does not couple with right handed fermions, it can be shown that the Greisen-Zatsepin-Kuzmin (GZK) effect would not be as strong, and the UHECR could propagate from sources 150 times further away than traditionally accepted (Tipler, 2005). Note that this does not mean the GZK effect would be nonexistent, rather, that it would be not as prominent; and indeed, it was not seen immediately in scientific investigations. Because of the theorized SU(2) dominant CBR, it can also be shown that the Sunyaev-Zel'dovich (SZ) effect should also be a factor of two too low when measured (Tipler, 2005; Tipler and Piasecki, 2018). 

Tipler is not the only researcher who theorizes about an SU(2) dominant CBR photon gas in order to solve cosmological problems. For example, Hofmann (2013) uses the idea of an SU(2) dominant CBR to explain the anomalous temperature anisotropies of the CBR at large angles. Hahn, Hofmann, and Kramer (2019) and Hofmann (2020) demonstrate various properties of an SU(2) dominant CBR cosmology, and how it fits current cosmological observations. 

Whereas a traditional CBR has a mean-free path of 3 Mpc for UHECR protons (Longair 1994 p. 340), Tipler shows that an SU(2) dominant CBR would have a mean free path of 450 Mpc (approximate redshift z = 0.1). Since the fractional energy loss by pion creation is approximately 10\%, this indicates UHECR protons might potentially be able to propogate from truly cosmic distances of up to 4.5 Gpc in this SU(2) cosmology (Tipler, 2005).

In this paper, I test how successful finding potential sources for the Northern Hemisphere UHECR is if the UHECR come from up to a distance of 450 Mpc (approximate redshift z = 0.1). I see if the theory is successful at identifying a significant number of possible celestial origin sources and compare it to traditional theories which assume the CBR photons have a strongly present U(1) component (i.e., a mean-free path of 3 Mpc). For the purposes of this study, I confine my analysis to finding potential sources from within the mean-free path distance. It is possible in the SU(2) CBR model for UHECR to come from distances much greater than z = 0.1, but for the sake of simplicity and demonstrating how well this theory can explain cosmological observations, I will assume a distance of z = 0.1 or less. 

The data I will be analyzing comes from the Northern Hemisphere High Resolution (HiRes) Fly's Eye Experiment. A separate paper analyzes Southern Hemisphere UHECR data available (Tipler and Piasecki, 2018), with impressive results (86\% identification rate for redshift z $\leq 0.1$ and within three degrees of the arrival direction). The HiRes data was collected from a project consisting of two telescope sites (sites I and II) that have been recording cosmic rays from 1999 to 2005 (Sokolsky, 2010). It consists of information about arrival direction, date and time, and measured energy for the UHECR. The analysis will be broken into two data sets based on the two telescope sites.

\section{Methods}

In order to find suitable sources for the UHECR observed, we looked at Active Galactic Nuclei (AGN), Quasars (QSO), and Seyfert galaxies within z = 0.1. These astronomical bodies are believed to be energetic enough to accelerate protons, Iron (Fe) Nuclei, and other particles which are believed to make up the UHECR to the observed large energies (Sokolsky, 2010). 

We confined our search within three degrees of the arrival direction, since the trajectories of the UHECR could be bent due to extra-galactic magnetic fields in space. To identify potential sources, I relied on the VizieR Catalog (VizieR) to search within three degrees of the arrival direction and chose the AGN, QSO, or Seyfert galaxy with redshift z less than 0.1. With much searching and sorting the data available, I compiled the data tables presented in Appendix I. The best sources (with smallest redshift, smallest offset angle, and correct astronomical type) were picked.

\section{Analysis and Discussion}

There are a total of 378 UHECR observed by HiRes over 1999 to 2005. Of the 378, 5 could not be paired with potential sources, 8 were identified with Quasars, 109 were identified with AGN, 79 were identified with Seyfert 1, 91 were identified with Seyfert 2, 6 as a type of (unspecified) Seyfert, 9 as BL Lac Type (a type of AGN), and the remaining 71 were identified as likely AGN, but tentatively so. We can therefore \textit{definitely} identify (378 - 5 - 71)/378 = 0.7998, or about 80\% , as being sources for the UHECR within a redshift of z = 0.1 or less. The 5 unidentified UHECR, I believe, are either due to the incompleteness of the current VizieR catalog or might lie outside of our idealistically imposed boundary of z = 0.1. The 71 tentatively identified AGN sources are only labeled `tentative' since they appear in AGN and Quasar catalogs, but I was unable to independently cross reference and verify that they are truly AGN. They are very likely to be AGN, but with the absence of independent verification I decided to just label them `likely' for the sake of clarity. If the likely AGN are indeed AGN, then a solid 98.7\% UHECR sources can be identified for the HiRes data.

%A potential source of error in this analysis is that some measurement methods for the redshift (such as the photometric measure) quantify the redshift in increments of 0.1. This has been used in celestial mapping projects like the WISEA project (MILLIQUAS). Sources that are $0.1\pm0.05$ will be measured as 0.1, and so objects that are a little over 0.1 might have been included in the data tables of Appendix I. For photometrically measured redshifts, we only include one significant digit (`0.1') in the tables to distinguish between higher accuracy measurements of the redshift (which are generally listed with three significant figures). 

%I think it is likely that further measurements in the proper regions of the night sky will reveal the other missing sources for the UHECR within three degrees. It is possible also that if we look beyond our idealist cut off point of z = 0.1, we might find the potential UHECR sources there.

One way this study could be extended is to investigate the Northern Sky UHECR `Hotspot' seen from the Telescope Array (TA) data (for example, see Kawata \textit{et al}, 2019). I have attempted to locate the original TA data, but unfortunately after many attempts, I was unable. I hope that the original data will be placed online, in full, in the near future. It would be most interesting to investigate the TA `Hotspot' in light of the theory described here. 

If the CBR photons have a strong U(1) component, cosmic rays would propagate with a mean-free path of 3 Mpc. There are almost no celestial bodies within 3 Mpc that could produce all the UHECR we see. If we assume that the UHECR are arriving from beyond the mean-free path, say at a generous z = 0.01 (15 times greater), there is still no salvation for the traditional theory.  If we confine ourselves to z $\leq 0.01$, we can see that only 49 of the 378 UHECR can be associated with AGN, QSO, and Seyfert galaxies from the tables present in Appendix I. This is a small 12.96\%, highlighting the stark deficiency of standard theory. The difference between 12.96\% and 80\% (possibly up to 98.7\%) cannot be stressed enough. Note with the 80\% we have neglected to take into account that sources beyond z = 0.1 are allowed (up to cosmic distances). Since this analysis was conducted with this idealistic constraint, it is quite likely that all UHECR sources will be identified if cosmic distances are taken into consideration.  

The 80\% identification rate fits very well with the SU(2) dominant theory based in the SM of Particle Physics. In a separate article, Tipler and Piasecki (2018) have shown they can identify 86\% of all Southern Hemisphere UHECR observed (as opposed to the standard 20\% within z = 0.01). I predict the success percentages will be even higher as more observations are made in the appropriate regions. The rates of success are highly parsimonious with the SU(2) dominant CBR explanation for UHECR. 

There are other lines of evidence that are converging, which suggest the SU(2) theory has serious merit. If the SU(2) idea is correct, Tipler (2005) predicted that the SZ effect should be too small by a factor of two. The surprising prediction has recently been corroborated. WMAP has observed that the SZ effect must be smaller than previously anticipated (Lieu \textit{et al}, 2006; Diego and Patridge, 2010), %is et al appropriate itext citation
 and the PLANCK Collaboration has observed it being too small by a factor of two (PLANCK Collaboration, 2013). 

Hofmann (2013) shows that an SU(2) dominant CBR can explain the observed breakdown of the statistical isotropy of the CBR. Modeling with the SU(2) dominant CBR can be shown to be consistent with local cosmological observations (Hahn, Hofmann, and Kramer, 2019). Consequences for the radiation and dark sectors of the cosmological model are explored (Hofmann, 2020).

Perhaps best of all, a recent direct measurement experiment has been conducted that attempts to measure the CBR to determine once and for all if the U(1) field is suppressed in CBR photons. A CBR photon with suppressed U(1) field would have not been detected in past experiments since it would have been likely absorbed by the detector and re-emitted as a photon with regular U(1) component (Tipler, 2005). Past observations of the CBR would have not noticed that the photons are SU(2) dominant. This, therefore, would require a very special set up to glean this particular, subtle property of the CBR photons. The recent results (publication pending) are all consistent with a U(1) suppressed present day CBR instead of the traditional theory. 

These independent lines of evidence suggest that the structure of the CBR is radically different from what the consensus previously envisioned. This has important implications for a number of different areas, including the present ratios of matter to anti-matter, the creation of matter and anti-matter, dark energy and dark matter, CBR temperature anisotropy at large angles, among other important topics (see Tipler, 2005; Hofmann, 2013, 2020). The problem of UHECR origin might not be a problem at all given the high identification success rate in this paper for the Northern Hemisphere and for the Southern Hemisphere (as described in the accompanying paper).

\section*{Acknowledgements}
\begingroup
    \fontsize{9pt}{12pt}\selectfont %\fontsize{9pts}{12pts} changes from 12 pts to 9 pts text
I would like to thank Angelica ``Clover" Robichaud of the University of South Carolina for helping me remain motivated and on task throughout this research project. Portions of this project took long, tedious hours of compilation and sorting through various astronomical bodies. Our long conversations while I worked helped me remain focused and complete this in a timely manner! I am also grateful to her for helping me perform some preliminary data analysis using the programming language R. Thanks goes to Clover, Andrea Robichaud, a mutual friend, and the reviewers of this manuscript for pointing out enhancements, typos, and corrections. I would also like to thank Dr. Frank Tipler of Tulane University for his input and suggestions for this article. 
\endgroup

\newpage\appendix

I now give lists for all the 378 HiRes UHECR for which we have been able to identify Active Galactic Nuclei (AGN), Seyfert galaxies, or Quasars (QSO) as sources. Note that BL Lac Type are a type of AGN. In the Tables below, photometrically measured redshifts are designed by `0.1' (with one significant digit) instead of values with three significant digits or more. 
AGN* are likely AGN, as they appear in the AGN and QSO catalogs, but I was unable to independently verify each identity.

\begin{table*}[h!]
  \begin{center}
    \caption{HiRes UHECR Year 1999-2000 Site 1}
    \label{tab:table1}
    \begin{tabular}{l|cl|c|cr|r} % <-- Alignments: 1st column left, 2nd middle and 3rd right, with vertical lines in between
      \textbf{Date} & \textbf{Energy} & \textbf{Source} & \textbf{Type} & \textbf{Redshift} & \textbf{Angle}\\
       & EeV & & & z & Degrees\\
      \hline

12 01 1999 & 18.19 & MCG -01.60.021 & Linear AGN & 0.023 & 1.82\\

12 01 1999 & 10.98 & IAU 0252-003 & Seyfert 2 & 0.029 & 1.24\\

01 06 2000 & 11.46 & NPM1G+79.0012 & Seyfert 2 & 0.056 & 2.62\\

01 09 2000 & 13.23 & SWIFT J0444.0+2859 & Seyfert 1 & 0.022 & 2.99\\

01 09 2000 & 15.46 & IC 2758 & Linear AGN & 0.020 & 2.13\\

02 01 2000 & 11.92 & MCG+08-15-056 & Seyfert 1& 0.041 & 1.05\\

02 02 2000 & 12.30 & WISEA J050809.25+370215.5  &  AGN* &  0.1& 1.38\\

03 03 2000 & 12.81 & 1FGL J1419.7+7731 & AGN & 0.006 & 1.54\\

03 29 2000 & 15.86 & Arp 102B & Seyfert 1 & 0.024 & 2.71\\

04 01 2000 & 48.15 & NPM1G-07.0469 &  AGN &  0.053& 2.94\\

04 05 2000 & 16.45 & NGC 5664 & AGN & 0.015 & 1.81\\

05 03 2000 & 17.58 & LQAC 271+026 001 & QSO & 0.063 & 1.01\\

09 03 2000 & 12.22 & 3C 84 & Seyfert 2 & 0.018 & 2.10\\

10 02 2000 & 30.25 & UM 85 & Seyfert 2 & 0.041  & 1.27\\

10 20 2000 & 14.65 & 2MASX J19052592+4227398 & Seyfert 1 & 0.028 & 2.80\\

10 26 2000 & 11.19 & NGC 7678 &  AGN & 0.012 & 2.51\\

10 29 2000 & 53.77 & PADOVA 1 & Seyfert 1 & 0.009 & 2.49\\

11 20 2000 & 10.04 & WISEA J235604.34+583240.2 & AGN* & 0.1 & 2.70\\

11 22 2000 & 10.95 & IC 2211 & Seyfert 2  & 0.018  & 2.48\\

11 25 2000 & 55.77 & MARK 1419 & Linear AGN & 0.015 & 0.75\\

11 25 2000 & 12.42 & GSC 4908 01685 &  Seyfert 2 & 0.040 & 1.76\\

11 27 2000 & 20.72 & WISEA J192858.92+564742.5 & AGN*  & 0.1 & 0.87\\

11 27 2000 & 11.29 & UGC 3374 & Seyfert 1  & 0.021 & 2.96\\

11 27 2000 & 17.43 & NGC 5322 & Seyfert & 0.006 & 1.48\\

11 28 2000 & 14.86 & WISEA J050809.25+370215.5 &  AGN* & 0.1 & 1.29\\

12 01 2000 & 23.97 & WISEA J194951.32+321106.9 & AGN* & 0.1 & 2.00\\

12 23 2000 & 53.64 & LAMOST 79.886+43.19993 &  AGN* & 0.080 & 2.99\\

12 25 2000 & 44.09 & WISEA J072737.83-145927.9 & AGN*  & 0.1 & 1.38\\

12 28 2000 & 22.36 & 6dFGS g0742283-045045 &  AGN* & 0.063  & 2.19\\

12 30 2000 & 24.07 & SNUQSO0644+3546 & QSO & 0.077 & 1.48\\

12 31 2000 & 21.91 & WISEA J054034.00-113558.1 & AGN* & 0.1 & 0.91\\

    \end{tabular}
  \end{center}
\end{table*}

\begin{table*}[h!]
  \begin{center}
    \caption{HiRes UHECR Year 2001 Site 1}
    \label{tab:table2}
    \begin{tabular}{l|cl|c|cr|r} % <-- Alignments: 1st column left, 2nd middle and 3rd right, with vertical lines in between
      \textbf{Date} & \textbf{Energy} & \textbf{Source} & \textbf{Type} & \textbf{Redshift} & \textbf{Angle}\\
       & EeV & & & z & Degrees\\
      \hline

01 26 2001 & 28.22 & IAU 0507+074 & Seyfert 2 & 0.019 & 1.75\\

01 29 2001 & 13.06 & WISEA J045411.72+441056.5 & AGN* & 0.1 & 2.98\\

01 29 2001 & 22.23 & UGC 6192 & AGN & 0.007 & 1.38\\

01 31 2001 & 10.14 & NGC 3259 & Seyfert 1 & 0.006 & 1.27\\

01 31 2001 & 16.77 & NGC 3161 & AGN & 0.021 & 2.15 \\

02 20 2001 & 26.44 & NGC 4725 & Seyfert 2 & 0.004 & 2.24\\

03 14 2001 & 17.76 & IRAS 01118+8455 &  Seyfert 2 & 0.056 & 0.84\\

03 19 2001 & 62.31 & PGC 1233260 & Seyfert 2 & 0.076 & 1.36\\

03 19 2001 & 19.51 & IAU 1034+061 & Seyfert 2& 0.012 & 2.14\\

03 19 2001 & 13.26 & UGC 10695SE &  Linear AGN & 0.028 & 2.92\\

03 24 2001 & 42.79 & MCG 2-26-023 & Linear AGN & 0.031 & 2.07\\

03 25 2001 & 37.45 & SDSS J101246.24+101039.4 & Seyfert 2 & 0.069 & 0.97\\

03 25 2001 & 18.62 & NGC 4151 &  Seyfert 1& 0.003 & 1.96\\

03 25 2001 & 14.95 & UGC 6192 & AGN & 0.007 & 2.14\\

04 16 2001 & 13.93 & NGC 5596 &  Seyfert 2& 0.011 & 2.81\\

05 19 2001 & 11.96 & NPM 1G-07.0450 & AGN &0.072  & 2.38\\

05 19 2001 & 10.45 & SDSS J164107.63+224924.8 & Seyfert 2&  0.034& 2.41\\

05 20 2001 & & Source not identifiable & & & \\

09 19 2001 & 20.92 & RXJ0024.7+0820 &  Seyfert 1& 0.067 & 2.96\\

    \end{tabular}
  \end{center}
\end{table*}

\begin{table*}[h!]
  \begin{center}
    \caption{HiRes UHECR Year 2002 Site 1}
    \label{tab:table3}
    \begin{tabular}{l|cl|c|cr|r} % <-- Alignments: 1st column left, 2nd middle and 3rd right, with vertical lines in between
      \textbf{Date} & \textbf{Energy} & \textbf{Source} & \textbf{Type} & \textbf{Redshift} & \textbf{Angle}\\
       & EeV & & & z & Degrees\\
      \hline

05 05 2002 & 18.20 & SDSS J081324.01+542236.7 & Seyfert 1 & 0.043 & 1.67\\

06 11 2002 & 10.66 & PGC 2772955 & Seyfert 1 & 0.089 & 2.21\\

07 10 2002 & 13.75 & WISEA J193016.76+085201.5 & AGN* & 0.1 & 2.45\\

07 12 2002 & 68.95 & SWIFT J2117.5+5139 & Blazar AGN & 0.002 & 1.57\\

08 14 2002 & 13.65 & MCG+04-06-043 & Seyfert 1 & 0.033 & 2.40\\

10 09 2002 & 12.73 & MKN 885 &  Seyfert 1 & 0.025 & 2.81\\

10 12 2002 & 13.71 & LQAC 309+032 001 & AGN* &  0.084& 2.74\\

11 04 2002 & 11.92 & NGC 7479 & Seyfert 2 & 0.007 & 1.53\\

11 07 2002 & 27.45 & SBS 0911+472 & AGN & 0.027 & 2.19\\

12 04 2002 & 11.42 & NGC 2776 & AGN & 0.009 & 1.45\\

12 08 2002 & 10.98 & UGC 12282 & Seyfert 1 & 0.017 & 1.24\\

12 30 2002 & 15.27 & UGC 12138 & Seyfert 1 & 0.025 & 2.66\\

    \end{tabular}
  \end{center}
\end{table*}

\begin{table*}[h!]
  \begin{center}
    \caption{HiRes UHECR Year 2003 Site 1}
    \label{tab:table4}
    \begin{tabular}{l|cl|c|cr|r} % <-- Alignments: 1st column left, 2nd middle and 3rd right, with vertical lines in between
      \textbf{Date} & \textbf{Energy} & \textbf{Source} & \textbf{Type} & \textbf{Redshift} & \textbf{Angle}\\
       & EeV & & & z & Degrees\\
      \hline

01 01 2003 & 25.18 & LQAC 092+007 001 & AGN* & 0.055 & 2.10\\

01 04 2003 & 20.82 & MARK 1055 & AGN & 0.036 & 2.51\\

01 04 2003 & 28.43 & NGC 71 &  Seyfert 2& 0.022 & 2.54\\

01 04 2003 & 15.02 & UGC 6192 & AGN  & 0.007 & 2.35\\

01 07 2003 & 16.03 & 2MASX J08185772-2252364 & Seyfert 1 & 0.035 & 1.85\\

01 30 2003 & 10.91 & MARK 1210 & Seyfert 2 & 0.013 & 1.78\\

02 01 2003 & 17.19 & Source not identifiable &  &  & \\

02 06 2003 & 19.55 & NGC 5879 & AGN & 0.003 & 1.18\\

02 24 2003 & 22.29 & WISEA J050809.25+370215.5 & AGN* & 0.1 & 2.10\\

03 03 2003 & 11.81 & WISEA J052026.82+333531.5 & AGN* & 0.1 & 0.35\\

03 28 2003 & 15.07 & IRAS F09182-0750 & Seyfert 2 & 0.020 & 2.37\\

03 29 2003 & 26.32 & IAU 1034-273 &  Linear AGN & 0.009 & 0.76\\

04 02 2003 & 16.19 & Ark 479 & Seyfert 2 & 0.019 & 2.89\\

04 26 2003 & 37.14 & RX J2135.9+4728 & Seyfert 1& 0.025 & 2.56\\

04 28 2003 & 10.86 & MKN 403 & Seyfert 2 & 0.024  & 2.81\\

04 29 2003 & 26.17 & IAU 1122+546 & Seyfert 1& 0.021 & 2.14\\

05 25 2003 & 16.73 & NGC 2655 & Linear AGN & 0.005 & 2.72\\

05 26 2003 & 11.71 & PGC 1548855 & Seyfert 2 & 0.026 & 1.47\\

05 26 2003 & 38.98 & 8C 0149+710 & BL Lac Type & 0.022  & 1.38\\

05 27 2003 & 16.01 & IAU 1344+038 & Seyfert 2 & 0.023 & 1.97\\

05 27 2003 & 49.34 & NPM1G+08.0399 & Seyfert 2 & 0.031 & 1.76\\

06 25 2003 & 12.19 & IRAS F17138-1017 & AGN* & 0.017 & 2.74\\

07 27 2003 & 11.66 & IGR J03334+3718 & Seyfert 1  & 0.055 & 2.85\\

08 25 2003 & 13.96 & IC 1368 & Seyfert 2 & 0.013 & 2.66\\

08 30 2003 & 10.33 & WISEA J193902.05+702930.3 & AGN* & 0.1 & 1.21\\

09 28 2003 & 11.97 & 1ES 1927+654 & Seyfert 1 & 0.017 & 2.96\\

09 29 2003 & 36.02 & IRAS 17596+4221 & Seyfert 2 & 0.053 & 1.19\\

10 19 2003 & 16.71 & NGC 1266 & AGN & 0.007 & 1.62\\

10 20 2003 & 15.42 & 3C 66.0B & Seyfert 1 & 0.021 & 2.75\\

10 23 2003 & 21.41 & IAU 2159-321 & Seyfert 2 & 0.008 & 2.57\\

10 24 2003 & 17.45 & PGC 2513233 & BL Lac Type & 0.036 & 2.43\\

10 25 2003 & 16.41 & Zw 493.004 & AGN & 0.043 & 2.54\\

10 27 2003 & 19.68 & UGC 11630 & Seyfert 2& 0.012 & 2.30\\

11 19 2003 & 68.06 & SDSS J221002.03+241228.0 & AGN*  & 0.087 & 2.62\\

11 20 2003 & 10.05 & UGC 524 & Seyfert 1 & 0.036 & 2.81\\

11 25 2003 & 11.55 & 4C 69.26 & Seyfert 1 & 0.097 & 0.86\\

12 01 2003 & 14.47 & MARK 703 & AGN  & 0.013 & 2.62\\

12 02 2003 & 57.40 & IRAS 06483-1955 & Seyfert 1 & 0.026 & 2.63\\

12 02 2003 & 36.19 & NGC 4203 &  Linear AGN & 0.004 & 0.86\\

12 15 2003 & 16.42 & NGC 2403 & AGN & 0.0004 & 2.78\\

12 16 2003 & 29.94 & NGC 1071 & Seyfert 2 & 0.038 & 2.89\\

12 18 2003 & 12.13 & KUG 0312+013 & Seyfert 2  & 0.023 & 0.96\\

12 21 2003 & 30.78 & MARK 341 & AGN & 0.017 & 2.99\\

12 22 2003 & 11.68 & PGC 2743419 & Seyfert 1 & 0.025 & 1.63\\

    \end{tabular}
  \end{center}
\end{table*}

\begin{table*}[h!]
  \begin{center}
    \caption{HiRes UHECR Year 2004 Site 1}
    \label{tab:table5}
    \begin{tabular}{l|cl|c|cr|r} % <-- Alignments: 1st column left, 2nd middle and 3rd right, with vertical lines in between
      \textbf{Date} & \textbf{Energy} & \textbf{Source} & \textbf{Type} & \textbf{Redshift} & \textbf{Angle}\\
       & EeV & & & z & Degrees\\
      \hline

02 18 2004 & 17.94 & MCG 9-11-023 & AGN  & 0.037 & 2.10\\

02 25 2004 & 22.12 & NGC 3858 & Seyfert 2 & 0.019 & 0.63\\

03 13 2004 & 11.44 & IRAS 17080+1347 & AGN* & 0.031 & 2.51\\

03 18 2004 & 19.68 & SDSS J155224.87+041726.2 & QSO & 0.046 & 1.21\\

03 20 2004 & 11.03 & NPM1G+08.0399 & Seyfert 2 & 0.031 & 2.73\\

03 21 2004 & 15.72 & NPM1G+41.0135 &  QSO & 0.022 & 2.56\\

03 22 2004 & 14.27 & IC 2226 & Linear AGN & 0.036 & 1.71\\

03 25 2004 & 31.56 & Messier 061 & Seyfert 2 & 0.005 & 1.19\\

04 12 2004 & 11.67 & RGB J0153+712 & BL Lac Type & 0.022 & 0.66\\

04 14 2004 & 14.49 & Zw 142.019 & AGN & 0.018 & 2.25\\

05 18 2004 & 41.42 & UGC 3995 & Seyfert 2  & 0.015 & 1.51\\

06 17 2004 & 17.62 & 2MASX J16375207+0016236 & Seyfert 1 & 0.047 & 2.99\\

06 20 2004 & 15.13 & Z 427-028 & Seyfert 2 & 0.031 & 2.99\\

07 12 2004 & 10.51 & IAU 1158+62 &  Linear AGN & 0.005 & 2.83\\

07 26 2004 & 11.36 & SDSS J201900.88+135204.9 & AGN*  & 0.1  & 1.72\\

08 13 2004 & 17.12 & KUG 2346+019A & AGN & 0.031 & 2.63\\

08 20 2004 & 46.56 & NGC7177 & Linear AGN & 0.004 & 2.65\\

08 22 2004 & 13.85 & PGC 2632696 & Blazar AGN & 0.050 & 2.80\\

09 09 2004 & 26.11 & SDSS J171725.53+291107.9 & AGN  & 0.029 & 1.13\\

09 11 2004 & 12.40 & KUG 1021+675 & Seyfert 2 & 0.039& 1.71\\

09 14 2004 & 40.50 & WISEA J162110.24-130409.9 & BL Lac Type & 0.1 & 1.07\\

09 15 2004 & 31.60 & NPM1G+24.0470 & Seyfert 1 & 0.039 & 1.19\\

09 15 2004 & 11.67 & NVSS J17586-0303 & QSO & 0.088 & 1.89\\

09 15 2004 & 126.72 & WISEA J065615.57-025247.4 &  AGN*& 0.1 & 2.29\\

09 21 2004 & 27.80 & 2MASX J21362534+1233333 & Linear AGN & 0.084 & 2.93\\

09 21 2004 & 14.42 & IRAS20351+2521 & AGN* & 0.034 & 2.81\\

10 09 2004 & 11.93 & UGC 11700 & Seyfert 2 & 0.017 & 0.78\\

10 12 2004 & 12.17 & KUG 0239+013 & AGN & 0.010 & 2.86\\

10 13 2004 & 11.78 & SWIFT J2010.7+4801 & Seyfert 2 & 0.025 & 2.94\\

10 13 2004 & 12.95 & LAMOST 81.464+55.07420 & AGN* & 0.036 & 1.61\\

10 13 2004 & 52.33 & IRAS 04361-1430 & Seyfert 2 & 0.035 & 2.26\\

10 14 2004 & 16.36 & IRAS 19370-0131 & AGN & 0.019 & 1.87\\

10 15 2004 & 19.84 & 3C 390.3 &  Seyfert 1 & 0.056 & 2.56\\

10 15 2004 & 50.17 & IC 164 &  Linear AGN & 0.017 & 2.88\\

11 14 2004 & 11.60 & WISEA J225013.81+685548.8 & AGN* &0.1 & 1.62\\

11 20 2004 & 19.66 & NGC 5033 & Seyfert 1 & 0.003 & 2.99\\

12 11 2004 & 17.57 & PGC 3095715 & Seyfert 1 & 0.018 & 2.29\\

12 12 2004 & 21.47 & LQAC 087+041 001 & AGN* & 0.045 & 2.87\\

12 15 2004 & 39.85 & ESO 548-G81 & Seyfert 1  & 0.014 & 2.90\\

12 16 2004 & 21.52 & LAMOST 92.706+33.93306 & AGN* & 0.009 & 0.73\\

12 16 2004 & 10.69 & IRAS 04061+1522 & Seyfert 1 & 0.045 & 0.31\\

12 18 2004 & 10.69 & SDSS J053156.04-062935.8 & AGN* & 0.1 & 0.94\\

12 18 2004 & 10.81 & LQAC 059+042 001 & AGN & 0.066 & 1.08\\

12 19 2004 & 15.13 & MARK 477 & Seyfert 2 & 0.038 & 0.65\\

    \end{tabular}
  \end{center}
\end{table*}

\begin{table*}[h!]
  \begin{center}
    \caption{HiRes UHECR Year 2005 Site 1}
    \label{tab:table1}
    \begin{tabular}{l|cl|c|cr|r} % <-- Alignments: 1st column left, 2nd middle and 3rd right, with vertical lines in between
      \textbf{Date} & \textbf{Energy} & \textbf{Source} & \textbf{Type} & \textbf{Redshift} & \textbf{Angle}\\
       & EeV & & & z & Degrees\\
      \hline

01 02 2005 & 11.60 & Source not identifiable &  &  & \\ 

01 02 2005 & 22.87 & PGC 1036549 & AGN & 0.064 & 2.22\\

01 07 2005 & 18.34 & WISEA J022109.52+625357.2 & AGN* & 0.1 & 0.90\\

01 07 2005 & 19.35 & IRAS 04589+5922 & QSO & 0.096 & 2.62\\

01 13 2005 & 16.83 & UGC 6125 & Seyfert 2 & 0.021 & 1.47\\

01 13 2005 & 29.44 & MKN 885 & Seyfert 1  & 0.025 & 1.50\\

01 16 2005 & 10.36 & NGC 3822 & Seyfert 1 & 0.021 & 2.66\\

02 05 2005 & 11.42 & RXS J05361+8223 & AGN & 0.051 & 2.26\\

02 09 2005 & 10.04 & MKN 993 & Seyfert 1 & 0.015 & 2.07\\

03 05 2005 & 44.39 & MARK 348 & Seyfert 2 & 0.014 & 2.33\\

03 05 2005 & 12.23 & IC 907 & Seyfert 2 & 0.015 & 1.52\\

03 07 2005 & 13.46 & PGC 3095749 & Seyfert 1  & 0.066 & 2.54\\

03 08 2005 & 10.36 & MCG 10-18-021 & Linear AGN & 0.044 & 1.79\\

03 08 2005 & 16.13 & NGC 3855 & Linear AGN & 0.031 & 1.93\\

03 09 2005 & 19.53 & MCG 12-15-013 & AGN & 0.025 & 2.45\\

03 11 2005 & 10.06 & KUG 1618+402 & Seyfert 1 & 0.028 & 2.01\\

03 12 2005 & 13.58 & NGC 6423 & Seyfert 1 & 0.024 & 1.82\\

03 31 2005 & 19.43 & LCRS B110045.0-063806 & Seyfert 1 & 0.025 & 2.56\\

03 31 2005 & 13.92 & LQAC 097+025 001 & AGN* & 0.083 & 2.64\\

04 01 2005 & 21.14 & NPM1G+81.0074 & AGN & 0.041 & 2.22\\

04 01 2005 & 71.49 & NGC 4968 & Seyfert 2 & 0.010 & 2.36\\

04 07 2005 & 11.74 & SWIFT J1933.9+3258 & QSO & 0.056 & 2.15\\

05 09 2005 & 19.08 & WISEA J200121.33+141536.9 & AGN* & 0.1 & 2.51\\

05 13 2005 & 10.26 & MCG 4-39-017 & Seyfert 2 & 0.031 & 2.16\\

05 29 2005 & 13.18 & IGR J20450+7530 & Seyfert 1  & 0.095 & 1.78\\

07 08 2005 & 27.96 & WISEA J190954.06+202129.0 & AGN* & 0.1 & 2.35\\

08 05 2005 & 24.70 & IRAS 21497-0824 & Seyfert 2 & 0.035 & 1.58\\

08 05 2005 & 14.80 & RX J18393+6544 & QSO & 0.082 & 0.56\\

08 31 2005 & 10.32 & IAU 1637+826 & Seyfert 2 & 0.024 & 1.11\\

09 01 2005 & 10.38 & WISEA J183813.45+260039.7 & AGN* & 0.1 & 1.61\\

09 01 2005 & 21.26 & 4C +47.63 & AGN*  & 0.045 & 1.73\\

09 02 2005 & 11.65 & 1RXS J050258.5+225949 & Seyfert & 0.057 & 0.98\\

09 02 2005 & 15.76 & MKN 279 & Seyfert 1  & 0.031 & 2.41\\

10 09 2005 & 34.61 & NGC 2824 & AGN & 0.009 & 1.86\\

10 24 2005 & 14.56 & IAU 2304+042 & Seyfert 1 & 0.042 & 2.49\\

10 25 2005 & 68.48 & Z 229-015 & Seyfert 1 & 0.028 & 1.21\\

10 26 2005 & 10.86 & IAU 2359+030 & Seyfert 1 & 0.026 & 1.01\\

10 31 2005 & 10.84 & IC 2227 & Seyfert 2 & 0.032 & 2.98\\

11 02 2005 & 17.57 & IAU 0446-064 & Seyfert 2 & 0.015 & 0.96\\

11 04 2005 & 10.77 & PGC 1137099 & Linear AGN & 0.053 & 2.28\\

11 04 2005 & 11.18 & IGR J03334+3718 & Seyfert 1 & 0.055 & 0.93\\

11 05 2005 & 14.55 & IRAS 04154+1755 & Seyfert 2 & 0.056 & 2.42\\

11 10 2005 & 13.08 & NGC 3735 & Seyfert 2 & 0.009 & 2.51\\

    \end{tabular}
  \end{center}
\end{table*}

\begin{table*}[h!]
  \begin{center}
    \caption{HiRes UHECR Year 1999-2000 Site 2}
    \label{tab:table1}
    \begin{tabular}{l|cl|c|cr|r} % <-- Alignments: 1st column left, 2nd middle and 3rd right, with vertical lines in between
      \textbf{Date} & \textbf{Energy} & \textbf{Source} & \textbf{Type} & \textbf{Redshift} & \textbf{Angle}\\
       & EeV & & & z & Degrees\\
      \hline
12 11 1999 & 11.20 & NGC 787 & Linear AGN & 0.016 & 2.97\\

01 04 2000 & 14.40 & UGC 3179 & AGN  & 0.028 & 0.38\\

01 15 2000 & 36.32 & PGC 1535850 & Linear AGN & 0.029 & 0.52\\

01 30 2000 & 12.63 & SDSS J104243.85+314121.8 & AGN & 0.035 & 2.41\\

02 02 2000 & 16.05 & Source not identifiable &  &   & \\

02 06 2000 & 40.84 & WISEA J062944.64+133201.5 & AGN* & 0.1 & 1.20\\

02 06 2000 & 30.28 & NGC 2624 & Linear AGN & 0.014 & 1.15\\

02 29 2000 & 18.35 & IC 907 & Seyfert 2  & 0.015 & 1.55\\

03 06 2000 & 15.58 & PGC 2790281 & Seyfert 1 & 0.081 & 0.72\\

03 06 2000 & 24.92 & GB6 J0601+5315 & BL Lac Type & 0.052 & 2.58\\

03 29 2000 & 12.33 & MCG 6-31-092 & Linear AGN & 0.043 & 2.18\\

03 31 2000 & 14.51 & NGC 2787 & Linear AGN  & 0.002 & 2.04\\

04 04 2000 & 16.74 & TEX 0554+534 & BL Lac Type & 0.036 & 2.06\\

04 06 2000 & 17.05 & IGR J20187+4041 & AGN & 0.014 & 2.89\\

04 09 2000 & 53.34 & UGC 5771 & Linear AGN  & 0.025 & 1.81\\

05 03 2000 & 40.79 & IGR J21277+5656 & Seyfert 1 & 0.015 & 1.79\\

09 04 2000 & 38.05 & WISEA J211823.95+230802.5 &  AGN* & 0.1 & 0.23\\

10 05 2000 & 12.91 & PGC 983528 & AGN & 0.074 & 2.92\\

10 07 2000 & 10.12 & PGC 3095729 & Seyfert 1 & 0.058 & 2.46\\

11 24 2000 & 28.03 & SDSS J231840.62+180204.9 &  AGN & 0.088 & 2.18\\

11 29 2000 & 12.53 & WISEA J210023.13+343521.5 & AGN*  & 0.1 & 1.49\\

12 01 2000 & 14.47 & PGC 3095730 & Seyfert 1 & 0.033 & 1.28\\

12 02 2000 & 11.94 & IRAS05129+5128 & AGN*  & 0.028 & 1.94\\

12 05 2000 & 19.49 & NGC 3735 & Seyfert 2 & 0.009 & 2.19\\

12 21 2000 & 29.69 & WISEA J040521.96+321900.6 & AGN* & 0.1 & 2.39\\

12 23 2000 & 48.59 & NGC 2841 & Linear AGN  & 0.002 & 1.95\\

    \end{tabular}
  \end{center}
\end{table*}

\begin{table*}[h!]
  \begin{center}
    \caption{HiRes UHECR Year 2001 Site 2}
    \label{tab:table1}
    \begin{tabular}{l|cl|c|cr|r} % <-- Alignments: 1st column left, 2nd middle and 3rd right, with vertical lines in between
      \textbf{Date} & \textbf{Energy} & \textbf{Source} & \textbf{Type} & \textbf{Redshift} & \textbf{Angle}\\
       & EeV & & & z & Degrees\\
      \hline

01 26 2001 & 19.15 & IRAS 08054+6824 & Seyfert 2 & 0.041 & 2.71\\

01 26 2001 & 12.79 & NGC 3561B & Linear AGN & 0.028 & 2.18\\

01 29 2001 & 26.00 & IAU 1043+140 & Seyfert 1 & 0.010 & 1.85\\

02 18 2001 & 105.71 & NPM1G+23.0202 & Seyfert 1  & 0.021 & 0.28\\

02 26 2001 & 24.29 & UGC 8621 & Seyfert 2 & 0.020 & 2.97\\

03 20 2001 & 16.87 & WISEA J042036.35+112641.8 & AGN* & 0.1 & 0.59\\

03 23 2001 & 17.43 & IAU 1353+407 & Linear AGN & 0.009 & 2.49\\

04 25 2001 & 10.42 & NGC 6434 & Seyfert 1 & 0.008 & 2.19\\

05 19 2001 & 18.33 & IRAS F16399-0937 & Linear AGN & 0.027 & 2.67\\

05 22 2001 & 17.90 & PGC 1143766 & Seyfert 2 & 0.049 & 0.93\\

05 23 2001 & 21.00 & NVSS J18564-2346& AGN & 0.057 & 2.68\\

05 23 2001 & 10.58 & MARK 1361 & Seyfert 2 & 0.023 & 2.67\\

07 18 2001 & 12.57 & MKN 357 & Seyfert & 0.053 & 0.56\\

07 18 2001 & 13.99 & IGR J02504+5443 & Seyfert 2 & 0.015 & 1.34\\

07 21 2001 & 24.02 & Source not identifiable &  &  & \\

07 27 2001 & 13.94 & MKN 359 & Seyfert 1 & 0.017 & 3.00\\

09 20 2001 & 10.70 & 1ES 1927+654 & Seyfert 1 & 0.017 & 1.04\\

09 21 2001 & 43.84 & NGC 1218 & Seyfert 1 & 0.029 & 2.35\\

09 22 2001 & 25.10 & UGC 10713 & AGN  & 0.004 & 2.98\\

    \end{tabular}
  \end{center}
\end{table*}

\begin{table*}[h!]
  \begin{center}
    \caption{HiRes UHECR Year 2002 Site 2}
    \label{tab:table1}
    \begin{tabular}{l|cl|c|cr|r} % <-- Alignments: 1st column left, 2nd middle and 3rd right, with vertical lines in between
      \textbf{Date} & \textbf{Energy} & \textbf{Source} & \textbf{Type} & \textbf{Redshift} & \textbf{Angle}\\
       & EeV & & & z & Degrees\\
      \hline

06 06 2002 & 14.14 & MKN 417 & Seyfert 2 & 0.033 & 2.21\\

07 08 2002 & 22.74 & IC 1368 & Seyfert 2 & 0.013 & 0.64\\

07 11 2002 & 13.52 & IRAS F20550+1655 & AGN* & 0.036 & 0.61\\

07 12 2002 & 17.79 & MKN 291 & Seyfert 1 & 0.035 & 2.64\\

07 13 2002 & 13.71 & NPM1G+44.0256 & Seyfert 1 & 0.055 & 1.99\\

08 09 2002 & 18.44 & SDSS J21167+1102 & AGN & 0.081 & 2.60\\

09 09 2002 & 19.32 & IAU 1701+315 & Linear AGN & 0.034 & 2.45\\

09 13 2002 & 11.73 & WISEA J201919.03+192452.3 & AGN* & 0.1 & 1.17\\

09 14 2002 & 11.22 & IAU 2325+085 & Seyfert 2 & 0.030 & 2.95\\

10 09 2002 & 11.42 & Mrk 6 & Seyfert 1 & 0.019 & 1.20\\

10 12 2002 & 54.84 & WISEA J194254.05+165249.4 & AGN* & 0.1 & 1.20\\

10 14 2002 & 13.37 & IAU 0645+744 &  Seyfert 1 & 0.0185 & 2.91\\

11 03 2002 & 16.32 & PGC 3081156 & AGN & 0.039 & 1.40\\

11 12 2002 & 10.07 & IRAS F00506+7248 & AGN* & 0.016 & 1.19\\

12 03 2002 & 10.74 & WISEA J064047.67+180334.0 & AGN* & 0.1 & 1.24\\ 
    \end{tabular}
  \end{center}
\end{table*}

\begin{table*}[h!]
  \begin{center}
    \caption{HiRes UHECR Year 2003 Site 2}
    \label{tab:table1}
    \begin{tabular}{l|cl|c|cr|r} % <-- Alignments: 1st column left, 2nd middle and 3rd right, with vertical lines in between
      \textbf{Date} & \textbf{Energy} & \textbf{Source} & \textbf{Type} & \textbf{Redshift} & \textbf{Angle}\\
       & EeV & & & z & Degrees\\
      \hline

01 04 2003 & 12.61 & WISEA J022109.52+625357.2 & AGN* & 0.1 & 1.31\\

01 05 2003 & 14.89 & WISEA J000019.48+621524.1	 & AGN* & 0.1 & 1.20\\

01 05 2003 & 18.85 & NGC 3147 & Seyfert 2 & 0.01 & 2.95\\

01 06 2003 & 62.84 & RX J03140+2445 & Seyfert & 0.056 & 1.43\\

01 29 2003 & 11.77 & PGC 1596882 & Seyfert 2 & 0.030 & 1.26\\

02 01 2003 & 31.76 & UGC 3752 & Seyfert 2 & 0.016 & 2.10\\

02 01 2003 & 111.09 & NGC 5940 & Seyfert 1 & 0.034 & 0.96\\

02 03 2003 & 20.23 & IAU 1219+047 & Seyfert 2 & 0.005 & 1.56\\

02 07 2003 & 12.45 & NGC 1961 & Linear AGN & 0.012 & 1.78\\

02 07 2003 & 52.16 & IAU 0931+103 & Linear AGN & 0.011 & 1.92\\

02 23 2003 & 10.59 & IAU 1426+276 & Seyfert & 0.014 & 2.36\\

03 25 2003 & 13.32 & UGC 10774 & Linear AGN & 0.030 & 1.71\\

03 29 2003 & 26.98 & MARK 287 & AGN &  0.038 & 2.35\\

03 30 2003 & 25.71 & SDSS J094328.73+524811.4 &  Seyfert 1 & 0.047 & 1.28\\

03 30 2003 & 10.10 & PGC 1388003 & Linear AGN & 0.022 & 2.09\\

03 30 2003 & 16.87 & 2MASX J18263239+3251300 & Seyfert 2 & 0.022 & 1.39\\

03 31 2003 & 15.09 & NGC 5631 & Seyfert 2 & 0.007 & 1.55\\

04 02 2003 & 11.85 & NGC 5674 & Seyfert 1 & 0.025 & 2.29\\

04 28 2003 & 20.34 & MCG -1-40-001 & Seyfert 2 & 0.023 & 1.91\\

05 03 2003 & 38.23 & NGC 3392 & AGN & 0.011 & 1.81\\

05 25 2003 & 34.05 & NVSS J19551-0338 & AGN & 0.056 & 2.11\\

05 26 2003 & 12.53 & PGC 1414609 & Seyfert 2 & 0.050 & 1.19\\

06 02 2003 & 15.44 & LQAC 253-007 001 & AGN* &0.069& 1.98\\

06 26 2003 & 34.70 & WISEA J202433.08+283456.1  & AGN* & 0.1 & 1.18\\

07 03 2003 & 85.62 & PGC 1438766 & AGN & 0.033 & 1.27\\

08 26 2003 & 20.59 & LQAC 060+034 001 & AGN* & 0.078  & 1.13\\

08 31 2003 & 17.20 & MKN 509 & Seyfert 1 & 0.036 & 2.49\\

09 01 2003 & 14.33 & 3C 449.0 & AGN & 0.017 & 2.46\\

09 22 2003 & 34.03 & WISEA J230022.97+531500.9 & AGN* & 0.1 & 0.47\\

09 22 2003 & 10.93 & MCG +09-13-070 & AGN & 0.019 & 2.58\\

09 23 2003 & 19.87 & SWIFT J02163+5128 & Seyfert 2 & 0.029 & 1.78\\

09 26 2003 & 11.30 & NPM1G+48.0362 & Seyfert 1 & 0.054 & 2.26\\

09 28 2003 & 46.35 & PGC 2790281 & Seyfert 1 & 0.081 & 2.92\\

09 29 2003 & 11.80 & LQAC 311+057 004 & AGN* & 0.038  & 0.37\\

10 01 2003 & 12.11 & RXS J22197+2120 & BL Lac Type & 0.020 & 1.34\\

10 22 2003 & 11.79 & SWIFT J0036.3+4540 & Seyfert 1 & 0.047 & 2.75\\

10 23 2003 & 11.20 & LEDA 67084 & Seyfert 1 & 0.025 & 2.45\\

10 23 2003 & 12.81 & UGC 3157 & Seyfert 2 & 0.015 & 2.45\\

10 27 2003 & 16.24 & IRAS 00065+1411 & Seyfert 2 & 0.040 & 2.68\\

10 28 2003 & 31.87 & WISEA J030908.83+100758.6 & AGN* & 0.1 & 0.45 \\

10 28 2003 & 50.33 & IRAS05129+5128 & AGN* & 0.028 & 2.79\\

10 29 2003 & 12.15 & NVSS J20045-0014 & AGN & 0.085 & 0.80\\

10 29 2003 & 10.08 & NGC 2782 & Seyfert 2 & 0.008 & 2.49\\

11 19 2003 & 11.73 & NGC 3079 & Seyfert 2 & 0.004 & 1.35\\

11 20 2003 & 13.35 & WISEA J003925.21+710719.0 & AGN* & 0.1 & 2.51\\

11 20 2003 & 14.64 & IGR J02504+5443 & Seyfert 2 & 0.015 & 1.51\\

11 21 2003 & 12.34 & IRAS 01118+8455 & Seyfert 2 & 0.056 & 2.45\\

11 25 2003 & 14.26 & WISEA J001833.69+552151.0 &  AGN* & 0.1 & 2.61\\

11 28 2003 & 11.49 & WISEA J225818.06+451634.7 & AGN* & 0.1 & 1.58\\

    \end{tabular}
  \end{center}
\end{table*}

\begin{table*}[h!]
  \begin{center}
    \caption{HiRes UHECR Year 2004 Site 2}
    \label{tab:table1}
    \begin{tabular}{l|cl|c|cr|r} % <-- Alignments: 1st column left, 2nd middle and 3rd right, with vertical lines in between
      \textbf{Date} & \textbf{Energy} & \textbf{Source} & \textbf{Type} & \textbf{Redshift} & \textbf{Angle}\\
       & EeV & & & z & Degrees\\
      \hline

02 14 2004 & 48.69 & IRAS05442+1732 & AGN* & 0.018 & 2.34\\

02 15 2004 & 22.31 & MKN 1 & Seyfert 2 & 0.016 & 1.50\\

02 15 2004 & 58.93 & NGC 3032 & AGN & 0.005 & 2.14\\

02 15 2004 & 14.67 & NGC 777 &  Seyfert 2  & 0.017 & 1.72\\

02 15 2004 & 16.06 & ZW VII 31 & AGN* & 0.054 & 2.01\\

02 15 2004 & 37.73 & ESO 442-G20 & AGN & 0.055 & 1.86\\

02 16 2004 & 12.72 & 3C 66.0B & Seyfert 1 & 0.021 & 2.88\\

02 25 2004 & 38.61 & NGC 5899 & Seyfert 2 & 0.008 & 2.70\\

02 25 2004 & 11.44 & LCRS B114409.4-032545 & AGN & 0.049 & 2.33\\

03 15 2004 & 10.84 & NGC 3177 & AGN & 0.004 & 2.99\\

03 17 2004 & 23.98 & WISEA J070444.29-111804.5 & AGN* & 0.1 & 0.44\\

03 17 2004 & 18.81 & WAS 96 &  AGN & 0.034 & 2.20\\

03 17 2004 & 15.29 & ESO 576-G77 & Linear AGN & 0.018 & 2.89\\

03 18 2004 & 20.51 & IAU 1548-037 & Seyfert 2 & 0.030 & 2.73\\

03 20 2004 & 29.25 & MKN 732 & Seyfert 1 & 0.030 & 2.76\\

04 13 2004 & 11.70 & SDSS J134632.14+642325.1  & Seyfert 2 & 0.024 & 1.58\\

04 13 2004 & 26.94 & RN 73 & Seyfert 1 & 0.049 & 0.96\\

04 20 2004 & 22.10 & MCG+09-13-070 & AGN & 0.019 & 2.62\\

04 23 2004 & 12.48 & IAU 1615+061 & Seyfert 1 & 0.038& 2.79\\

05 17 2004 & 17.59 & SDSS J204224.74+144455.3 &  AGN* & 0.1 & 1.51\\

06 13 2004 & 12.62 & WISEA J194254.05+165249.4 & AGN* & 0.1 & 1.80\\

06 14 2004 & 31.13 & NGC 4939 & Seyfert 2 & 0.010 & 1.67\\

06 17 2004 & 11.82 & IRAS F17138-1017 & AGN & 0.017 & 2.50\\

06 17 2004 & 13.08 & WISEA J180316.39+044327.4 & AGN* & 0.1 &1.03 \\

06 20 2004 & 40.19 & NGC 3738 & AGN & 0.001 & 1.02\\

07 12 2004 & 15.33 & 6dFGS g1607026-204304 & AGN* & 0.006  & 1.84\\ 

07 18 2004 & 13.66 & RXS J22197+2120 & BL Lac Type & 0.020 & 2.88\\

07 24 2004 & 26.40 & MKN 1 & Seyfert 2  & 0.016  & 1.86\\

08 12 2004 & 17.51 & MKN 845 &  Seyfert 1 & 0.046 & 2.35\\

09 09 2004 & 14.16 & KUG 2346+019A & AGN & 0.031 & 2.78\\

09 14 2004 & 28.90 & 5BZG J2346+4024& BL Lac Type  & 0.084 & 1.38\\

09 14 2004 & 71.07 & SDSS J041921.18+255304.6 & Seyfert 2 & 0.065 & 2.76\\

09 16 2004 & 19.59 & WISEA J192602.16+251357.6 & AGN* & 0.1 & 2.58\\

10 07 2004 & 28.06 & NGC 237 & AGN & 0.014 & 1.77\\

10 13 2004 & 12.28 & MCG -01-60-021 & Linear AGN & 0.023 & 0.40\\

11 06 2004 & 17.55 & IRAS 01118+8455 & Seyfert 2 & 0.056 & 2.96\\

12 03 2004 & 11.82 & NGC 926 & AGN & 0.021 & 1.87\\

12 11 2004 & 17.05 & NGC 3497 & Linear AGN & 0.012& 1.45\\

12 15 2004 & 23.31 & NGC 2638 & Linear AGN &  0.013& 1.45\\

12 15 2004 & 39.25 & WISEA J064622.49-120847.4 &  AGN*&  0.1& 2.26\\

12 16 2004 & 24.24 & IAU 0851+589 & Seyfert 2 & 0.003 & 2.01\\

12 16 2004 & 45.54 & IRAS 01118+8455 & Seyfert 2 & 0.056 & 2.46\\

12 19 2004 & 14.92 & AKN 202 & Seyfert 1  & 0.028 & 2.80\\

12 19 2004 & 27.00 & IRAS 06273+6858 & AGN & 0.065 & 2.93\\

12 19 2004 & 13.65 & NPM1G-18.0376 & AGN & 0.065 & 2.07\\

    \end{tabular}
  \end{center}
\end{table*}

\begin{table*}[h!]
  \begin{center}
    \caption{HiRes UHECR Year 2005 Site 2}
    \label{tab:table1}
    \begin{tabular}{l|cl|c|cr|r} % <-- Alignments: 1st column left, 2nd middle and 3rd right, with vertical lines in between
      \textbf{Date} & \textbf{Energy} & \textbf{Source} & \textbf{Type} & \textbf{Redshift} & \textbf{Angle}\\
       & EeV & & & z & Degrees\\
      \hline

01 07 2005 & 13.87 & NGC 1050 & Seyfert 2 & 0.013 & 2.57\\

01 08 2005 & 23.56 & IRAS05442+1732 & AGN* & 0.019 & 2.55\\

01 09 2005 & 24.12 & NGC 3035 & Seyfert 1 & 0.015 & 1.82\\

02 02 2005 & 15.11 & UGC 8107 & Linear AGN & 0.028 & 1.87\\

02 04 2005 & 16.18 & UGC 5771 & Linear AGN & 0.025 & 1.82\\

02 09 2005 & 16.47 & IC 632 & Seyfert 2 & 0.019 & 0.63\\

02 13 2005 & 38.30 & B3 0920+416 & AGN & 0.028 & 1.83\\

03 05 2005 & 13.53 & IGR J02504+5443 & Seyfert 2 & 0.015 & 1.79\\

03 06 2005 & 15.59 & IAU 1426+276 & Seyfert & 0.014 & 2.58\\

03 07 2005 & 22.81 & MCG 8-30-048 & Linear AGN & 0.0175 & 2.21\\

03 07 2005 & 16.79 & NGC 4102 & Linear AGN & 0.003 & 1.54\\

03 12 2005 & 36.10 & MRK 635 & Linear AGN  & 0.024 & 0.56\\

03 31 2005 & 28.66 & WISEA J235607.51+794402.0 & AGN  & 0.1 & 1.07\\

04 02 2005 & 12.55 & MCG -04.27.011 & AGN & 0.027 & 1.72\\

04 07 2005 & 37.66 & MCG -02.28.039 & Seyfert 1 & 0.026 & 2.92\\

07 06 2005 & 10.07 & WISEA J231003.43+565415.6 & AGN* & 0.1 & 1.55\\

07 07 2005 & 33.95 & NGC 5377 & AGN & 0.006 & 2.93\\

07 12 2005 & 17.80 & MCG +13-08-056 & Seyfert 2 & 0.037 & 2.78\\

08 31 2005 & 14.76 & 4C 69.26 & Seyfert 1 & 0.097 & 0.75\\

09 09 2005 & 20.96 & LEDA 67084 & Seyfert 1 & 0.025 & 2.18\\

09 26 2005 & 17.28 & IAU 1534+580 & Seyfert 1 & 0.030 & 1.29\\

09 26 2005 & 28.59 &IRAS F17544+4347 & AGN & 0.073 & 1.56\\

09 27 2005 & 13.39 & NGC 6786 & Seyfert 2 & 0.025 & 2.01\\

09 29 2005 & 15.83 & NGC 237 & AGN & 0.014 & 2.88\\

10 05 2005 & 18.78 & WISEA J225818.06+451634.7 & AGN*  & 0.1 & 0.80\\

10 06 2005 & 37.41 & IRAS07251-0248 & AGN* & 0.086 & 1.33\\

10 27 2005 & 10.44 & WISEA J063511.45+265706.8 & AGN* & 0.1 & 2.73\\

10 29 2005 & 11.93 & MARK 585 & AGN & 0.021 & 1.41\\

11 01 2005 & 11.15 & LQAC 319+073 001 & AGN* & 0.048 & 1.80\\

11 03 2005 & 26.61 & IAU 2329+286 & AGN & 0.019 & 1.09\\

11 06 2005 & 10.77 & PGC 1272062 & Seyfert 1 & 0.081 & 1.79\\

    \end{tabular}
  \end{center}
\end{table*}

%$\sqrt[3]{2}$

%$2/3$

%$$\displaystyle{\frac{2}{3}}$$
%This is a delta

%$\Delta$

%$\infty$

\end{document}